\documentclass[9pt,conference]{IEEEtran}
\usepackage{times}
\usepackage{url}
\usepackage{listings}
\usepackage{amsmath}
\usepackage[boxruled,linesnumbered]{algorithm2e}
\usepackage{algpseudocode}
\usepackage{tikz}
\usepackage{flushend}
\usepackage{array}

\usepackage{pgf}
\usetikzlibrary{positioning,shadows,arrows,automata,shapes}

\begin{document}

\pagenumbering{arabic}


\title{Learning Likely Invariants to Explain Why a Program Fails}


\author{
\IEEEauthorblockN{Jun Sun\IEEEauthorrefmark{1}, Long H. Pham\IEEEauthorrefmark{1},
Lyly Tran Thi\IEEEauthorrefmark{1}, Jingyi Wang\IEEEauthorrefmark{1}, Xin Peng\IEEEauthorrefmark{2}}
\IEEEauthorblockA{\IEEEauthorrefmark{1}ISTD Pillar, Singapore University of Technology and Design, Singapore}
\IEEEauthorblockA{\IEEEauthorrefmark{2}School of Computer Science and Shanghai Key Laboratory of Data Science, Fudan University, China}
}

\maketitle

\begin{abstract}
Debugging is difficult. Recent studies show that automatic bug localization techniques have limited usefulness. One of the reasons is that programmers typically have to understand why the program fails before fixing it. In this work, we aim to help programmers understand a bug by automatically generating likely invariants which are violated in the failed tests. Given a program with an initial assertion and at least one test case failing the assertion, we first generate random test cases, identify potential bug locations through bug localization, and then generate program state mutation based on active learning techniques to identify a predicate ``explaining'' the cause of the bug. The predicate is a classifier for the passed test cases and failed test cases. Our main contribution is the application of invariant learning for bug explanation, as well as a novel approach to overcome the problem of lack of test cases in practice. We apply our method to real-world bugs and show the generated invariants are often correlated to the actual bug fixes. 
\end{abstract}

\section{Introduction}
Debugging is an important part of software engineering and often considered to be difficult. Software engineering is the process of constructing a program based on a specification. The specifications, which asserts what is considered correct or otherwise buggy, may be missing in practice and may only exist in the programmer's mind. Ideally, if a specification that documents what is to be achieved for each statement is available, we can define a ``bug'' to be the first statement in the program where it fails to refine the specification. Debugging then can be done by contrasting the program against its specification to identify the first location where they differ. Without the specification, we are left with observations associated with the bug, e.g., which statements are executed in a failed test case; which statements are frequently executed in failed test cases; which conditions in the conditional statements are essential for reproducing the bug, etc. Based on these observations, extensive studies on bug localization have been conducted. Interested readers are referred  to~\cite{survey} for a survey of work prior to 2009 and~\cite{DBLP:conf/icst/MoonKKY14,DBLP:conf/issta/RobetalerFZO12} for some recent attempts. However the recent studies in~\cite{DBLP:conf/issta/ParninO11,icse2016} suggest that bug localization may not be sufficient as \emph{programmers have to understand the bug before fixing it}.

Inspired by their work, we propose a method to complement existing bug localization techniques in this work. 
We develop a software toolkit called \textsc{Ziyuan} to automatically generate likely invariants which are violated in the failed program execution. The goal is to help the programmers develop a high-level understanding of a bug.
Given a program (e.g., a Java method) with an assertion and at least one test case failing the assertion, \textsc{Ziyuan} first generates a set of test cases (by randomly instantiating the method parameters). Next, applying bug localization techniques~\cite{DBLP:conf/kbse/AbreuZG09}, \textsc{Ziyuan} identifies a list of ranked likely bug locations. \textsc{Ziyuan} then attempts to learn likely invariants for explaining the bug at these locations one-by-one. In particular, \textsc{Ziyuan} categorizes the program states at the location of all test cases into two sets, one containing the program states of those passed test cases and the other containing those of the failed test cases. Afterwards, \textsc{Ziyuan} employs machine learning techniques to learn a classifier between the two sets. Intuitively, the classifier is a likely invariant which explains the difference between the the passing test cases and the failing ones.

One essential problem of this approach is the lack of test cases, i.e., we might have only a very limited set of program states at a program location. In particular, if the likely bug location is in the middle of the program, it is in general hard to generate test cases to reach the location. As a result, the learned classifier is biased and may not be useful. To solve this problem, \textsc{Ziyuan} applies selective sampling~\cite{DBLP:conf/icml/OrabonaC11}, to iteratively generate ``artificial program states'' at the learning program location so as to learn a better classifier. That is, given a program location and a classifier for the program states (of the passed and failed test cases), we apply selective sampling to automatically compute the most informative program state for improving the classifier. \textsc{Ziyuan} then automatically mutates the program according to the computed program states, and re-runs the test cases. Based on the testing results, \textsc{Ziyuan} labels the program state accordingly (as either causing assertion failure or not) and refines the classifier. In this way, the classifier converges. We remark that these program states are artificial as they may not be reachable from the beginning of the program. Nonetheless, we show that the learned predicate correctly classifies program states at the program location and is useful in helping programmers understand the bug, as we show in the empirical studies. If we fail to find a classifier at a program location, \textsc{Ziyuan} takes another potential bug location and starts the same process from there. \textsc{Ziyuan} terminates when a predicate (i.e., a likely invariant) is identified or after exhausting the bug locations. The identified likely invariant is then presented to the user as a bug explanation.

To evaluate the effectiveness of \textsc{Ziyuan}, we apply \textsc{Ziyuan} to real-world bugs from open source projects and evaluate the generated bug explanations. Firstly, we show that the generated predicates are often correlated with the actual bug fixes. Then, we manually check whether the predicates always hold after the bug fixes or whether specific code is introduced in the fixed programs to handle the case when the predicate is not satisfied. We present detailed findings which suggest the usefulness of the generated predicates in bug comprehension. Secondly, as \textsc{Ziyuan} works by learning likely invariants, we compare \textsc{Ziyuan} with established invariant inference tools like Daikon~\cite{DBLP:conf/icse/ErnstCGN99} as well as FailureDoc~\cite{DBLP:conf/kbse/ZhangZE11} to show the difference. We further show, with examples, that \textsc{Ziyuan} complements existing bug localization techniques~\cite{survey}. Lastly, we conduct a user study by asking programmers to fix buggy programs with or without the help of \textsc{Ziyuan}. The result shows that the predicates generated by \textsc{Ziyuan} help bug understanding and fixing.

The rest of the paper is organized as follows. 
Section~\ref{details} presents the details of our approach using a running example. Section~\ref{experiments} presents the implementation of \textsc{Ziyuan} and the results of the empirical studies. Section~\ref{related} concludes with a review of related work. 

%
%
%
%
%

\begin{table*}[t]
\scriptsize
\centering
\caption{Test cases}
\begin{tabular}{|c|c|c|c|}
\hline
\textbf{test} & \textbf{Test Input Description} & \textbf{Pass/Fail} & \textbf{Ranked Features} \\
\hline
1 & 3 student objects with IDs 1,2,3 and scores 94, 60 and 100 & Fail & [94,94,60,100,1,0,2,0,3,0]\\
\hline
2 & 3 student objects with IDs 3,2,1 and scores 75, 90 and 80 & Pass & [90,75,90,80,3,0,2,0,1,0] \\
\hline
3 & 3 null student objects & Irrelevant & - \\
\hline
4 & 3 student objects with IDs 99,-10,0 and scores -33, 12 and 0 & Pass & [12,-33,12,0,99,0,-10,0,0,0] \\
\hline
\end{tabular}
\label{testcases}
\end{table*}

\section{Our Approach}\label{details}
\begin{figure}[t]
\centering
\includegraphics[width=9.5cm]{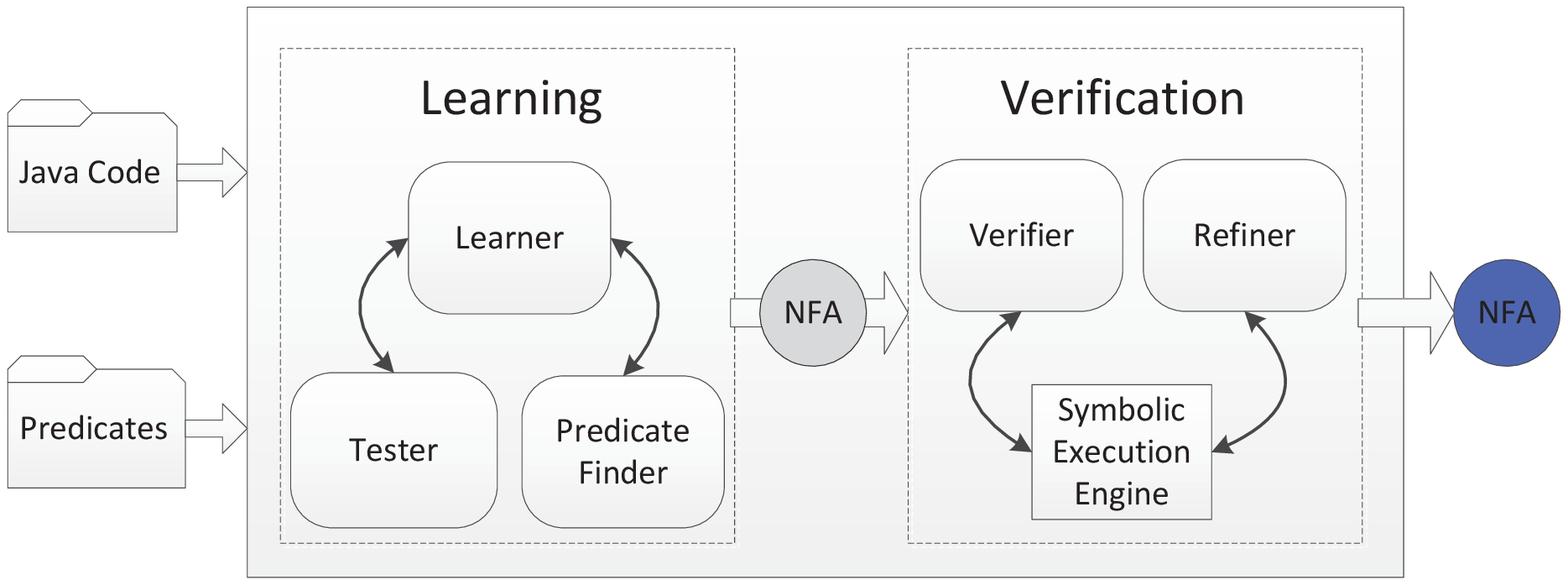}
\vspace{-8mm}
\caption{Overall workflow}
\label{workflow}
\end{figure}

In this section, we present details of our approach. We assume that the given program is deterministic, i.e., it is sequential and does not contain random number generation and there is no test harness problem. This assumption is necessary as our approach learns based on testing results. The overall workflow of \textsc{Ziyuan} is shown in Figure~\ref{workflow}. There are 6 steps, which are explained in sequence in the following.

The program shown in Figure~\ref{illustrative} is a toy example we designed to convey how \textsc{Ziyuan} works. The program contains a method $program$ which takes multiple objects of type $Stu$ (i.e., representing a student) as input and invokes method $standardize$ so as to standardize the students' scores through a crafted formula at line 9. Method $standardize$ takes an array of student objects and finds out the maximum score among the students and sets a new standardized score for each student in the array. We can \emph{manually} infer that the new score is always no more than 100, although it would be hard for program analysis methods like symbolic execution, due to the non-linear transformation.

A test case of a given program is a concretization of the parameters. Let us assume that a failed test case (test 1 in Table~\ref{testcases}) is given for the above program, with the input being three student objects with scores of 94, 60 and 100 respectively. The tester notices that the third student's new score is more than 100, which signals a bug in the code. Intuitively, this is because the last student object is missed when the maximum score $max$ is calculated, i.e., the bug manifests if the last student is the top scorer. The program can be fixed in different ways, e.g., at line 5 by changing the loop condition to $i < stus.length$, or at line 4 by setting $max$ to be the last student's score. 

In order to use \textsc{Ziyuan}, first the user is asked to provide an initial assertion \emph{in the program} based on the failed test case. For instance, we assume that the assertion at line 3 in Figure~\ref{illustrative} is added, which asserts that the third student's new score should not be more than 100. With this assertion, the failed test case results in assertion violation. 
We acknowledge the difficulty in writing assertions in general~\cite{DBLP:conf/fm/EstlerFNPM14} and remark that writing an assertion to capture the failure of a particular test case is often easier.
\begin{figure}[t]
{\scriptsize \begin{verbatim}
public static void program (Stu s1, Stu s2, Stu s3) {
1.  Stu[] list = new Stu[]{s1,s2,s3};
2.  standardize(list);	
3.  assert(s3.newscore <= 100);
}
private static void standardize(Stu[] stus) {
4.  int max = Integer.MIN_VALUE;
5.  for (int i = 0; i < stus.length-1; i++) {
6.     if (max < stus[i].score) {
7.	       max = stus[i].score;}
    }
    //version 1: max = 94;
    //version 2: max = 90;
    //version 3: max = 12;     	
8.  for (Stu stu: stus)
9.     stu.newscore = Math.sqrt((100-max)+stu.score)*10;
}
class Stu {
    int score; int ID; double newscore;
    public Stu (int s, int id) {score = s; ID = id;}
}
\end{verbatim}}
\vspace{-5mm}
\caption{An illustrative Java program}
\label{illustrative}
\end{figure}

\subsection{Step 1: Test Case Generation}
\textsc{Ziyuan} works better if a comprehensive set of test cases are provided. In practice, the set of user-provided test cases are often limited. Thus, in order to provide more initial data for bug localization as well as classification (as explained later), \textsc{Ziyuan} embeds an implementation of the Randoop algorithm~\cite{randoop} for random test case generation. That is, given a Java program, which is a method with multiple parameters, \textsc{Ziyuan} would generate arguments automatically for the method call (so as to construct test cases for the method). For each typed parameter, \textsc{Ziyuan} randomly generates a value from a pool of type-compatible values. This pool composes of a set of pre-defined values (e.g., random integers for an integer type, $null$ for a user-defined class) and type-compatible objects that have been generated during the testing process. In order to re-create the same object, we store the test case which produces the object. We refer the readers to the work in~\cite{randoop} for details on test case generation. We choose Randoop over other testing techniques because it is relatively (computationally) cheap. A systematic or more sophisticated testing method (e.g., dynamic symbolic execution~\cite{dart:PLDI2005,DBLP:journals/cacm/CadarS13} or genetic algorithm guided testing~\cite{DBLP:conf/qsic/FraserA11}) would possibly generate better test cases and improve \textsc{Ziyuan}'s performance. \\

\noindent \emph{Example} For the running example, let us assume three test cases are generated using random values as well as default values $null$ for all reference types, as shown in Table~\ref{testcases} (test 2, 3, and 4). In particular, test 2 does not trigger assertion violations. In test case 3, we assume that all three student objects are null. This could be the case since test cases are generated randomly. Executing test case 3 leads to an exception but not the assertion failure and we categorize it as irrelevant (i.e., it does not reach the assertion and we have no idea whether it would have satisfied it or not). Test case 4 has three student objects with unusual IDs and scores. This is possible as we do not have a specification on the range of scores and IDs.

\subsection{Step 2: Bug Localization}
The user-provided assertion can be considered as the very first bug explanation. It may not be informative though. In particular, it may not be intuitively associated to the cause of the bug if it is far away from the bug, i.e., whatever misbehavior the bug has caused may have been transformed out of shape through the subsequent statements. Therefore, in this step, we identify potential bug locations in the program so that we may generate bug explanations close to where the bug is in the code.

\textsc{Ziyuan} first applies a program slicer~\cite{jslicer} to identify the statements upon which the assertion has dependencies (including both control dependency and data dependency). In our running example, this includes all numbered statements. Next, we adopt existing bug localization techniques~\cite{survey} to offer clues on where the bug might be among those statements. In this work, we adopt Ochiai's approach~\cite{DBLP:conf/kbse/AbreuZG09}, which is an example of the spectrum based fault localization (SBFL) methods. In the following, we briefly introduce SBFL and refer the readers to~\cite{DBLP:conf/kbse/AbreuZG09,survey} for details.
SBFL techniques are designed based on the following intuitive idea: the more a statement is executed by the passed test cases, the less likely it is a bug; and the more it is executed by the failed test cases, the more likely it is. Given a set of passed test cases and failed test cases, SBFL computes a suspiciousness score for each statement in the program (based on how often it is executed by the passed/failed test cases). Different SBFL techniques use different functions to compute the suspiciousness. 
%
For instance, applying Ochiai's approach to our example with the four test cases, line 1,2,4 and 5 have the same suspiciousness 0.5, and line 3,6,7,8 and 9 have the same suspiciousness 0.57.
We refer the readers to~\cite{DBLP:conf/issta/ParninO11,DBLP:conf/icsm/LuciaLJB10} for an evaluation of the effectiveness of SBFL including Ochiai's approach. 

Recent empirical studies~\cite{DBLP:conf/issta/ParninO11,icse2016} suggest that existing bug localization techniques are not very accurate and have limited usefulness in practice. In our work, we do not assume that bug localization is precise. Rather, \textsc{Ziyuan} takes the suspicious program locations as input and attempts to generate a likely invariant at those locations one-by-one, and present the bug explanation to the users.

Furthermore, recall that we view a bug explanation as an inconsistency between the program behavior and its specification; we thus favor program locations where the program behavior can be naturally specified. For instance, if a statement in a loop has a high suspiciousness, \textsc{Ziyuan} would set out to look for a bug explanation after the loop because it is easier to specify the program's behavior there than in the middle of the loop\footnote{This avoids the loop invariant generation problem~\cite{DBLP:conf/cav/AlbarghouthiM13,DBLP:conf/cav/HassanBS12}.}. Furthermore, a block of sequential statements (without branching) often has the same suspiciousness, thus \textsc{Ziyuan} groups them and tries to generate only one bug explanation after the block.

In the case of our running example, among the likely bug location (i.e., all numbered lines), \textsc{Ziyuan} attempts to identify likely invariants at three program locations, i.e., right before line 8, or right after line 1, or right before line 5. Note that since line 3 is the assertion, \textsc{Ziyuan} ignores it since the initial assertion is already a good bug explanation there; line 5 is a part of the first loop and thus \textsc{Ziyuan} attempts to generate a bug explanation after the first loop (i.e., right before line 8); line 8 and 9 are a part of the second loop, which is followed the assertion and thus \textsc{Ziyuan} ignores them.

\subsection{Step 3: Feature Selection}
After step 2, a list of program locations have been identified. These program locations are ranked according to their suspiciousness score. Starting with the top program location in the list, we instrument the program and execute both the passed and failed test cases so as to collect the program states (i.e., valuation of all variables) at the location in all test cases.
In general, the number of variables accessible at a program location could be huge. 
\textsc{Ziyuan} uses the same program slicer to identify relevant ones (i.e., the variables which the assertion has a dependency on) and prunes the rest.

Next, \textsc{Ziyuan} categorizes the program states into two sets: $O^-$ containing those program states in the failed test cases and $O^+$ containing those in the passed test cases. Intuitively, there must be some difference between $O^-$ and $O^+$ which determines whether a test case fails or not. The question is what form of difference we should explore and how to identify them automatically. The answer to the first question is that we view a program state as a vector of features (in the form of float-type numbers) and a difference between the program states takes the form of a predicate on the features. The answer to the latter question is that we apply classification techniques from the machine learning community to identify such predicates.

In the following, we first show how to systematically obtain features from a program state. In general, there are both numerical-type (e.g., \textit{int}, \textit{boolean}) and categorical-type (e.g., \textit{Stu}) variables in Java programs. It is straightforward to cast the value of a numerical-type variable into a feature value. We need a systematic way of mapping a categorical-type object state to numerical values. 
Our approach is to systematically generate a \emph{numerical value graph} from each object type~\cite{tzuyu:ASE2013}.

\begin{figure}[t]
\begin{center}
{\scriptsize
\begin{tikzpicture}[
    reference/.style={rectangle, draw, rounded corners=1mm, text centered, anchor=north, minimum height=3mm, minimum width=8mm},
    leaf/.style={circle, draw, text centered, anchor=north},
    ->,>=stealth', shorten >=1pt, auto, initial text=, initial where=above, semithick,
    growth parent anchor=south, node distance=0.8cm
]
\node (students) [reference] {stus};
\node (rNull) [leaf,right=of students] {B};
\node (student1) [reference,below=of students] {stus[1]};
\node (rNullStu0) [leaf,left=of student1] {B};
\node (rNullStu1) [leaf,right=of student1] {B};
\node (student2) [reference,right=of rNullStu1] {...};
\node (student0) [reference,left=of rNullStu0] {stus[0]};
\node (length) [leaf,left=of student0] {I};
\node (id0) [leaf,below=of student0] {I};
\node (id1) [leaf,below=of student1] {I};
\node (score0) [leaf,left=of id0] {I};
\node (score1) [leaf,left=of id1] {I};
\node (newscore0) [leaf,right=of id0] {D};
\node (newscore1) [leaf,right=of id1] {D};
\node (id0) [leaf,right=of score0] {I};
\path[->] (students) edge node {$isNull$} (rNull);
\path[->] (students) edge node {} (student1);
\path[->] (students) edge node {} (student2);
\path[->] (students) edge node {} (student0);
\path[->] (students) edge[bend right=25] node {$length$} (length);
\path[->] (student1) edge node {$isNull$} (rNullStu1);
\path[->] (student1) edge[bend right=25] node {$score$} (score1);
\path[->] (student1) edge node {$ID$} (id1);
\path[->] (student1) edge node {$newscore$} (newscore1);
\path[->] (student0) edge node {$isNull$} (rNullStu0);
\path[->] (student0) edge[bend right=25] node {$score$} (score0);
\path[->] (student0) edge node {$ID$} (id0);
\path[->] (student0) edge node {$newscore$} (newscore0);
\end{tikzpicture}}
\end{center}
\vspace{-5mm}
\caption{The numerical value graph for object $stus$}
\label{tree}
\end{figure}
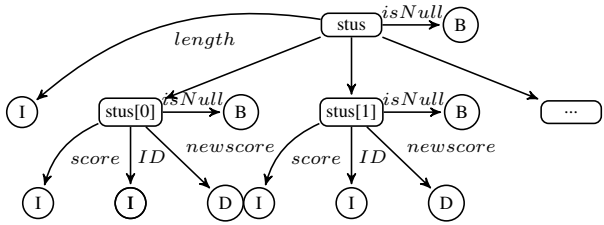

We illustrate how to construct numerical value graphes using an example. Figure~\ref{tree} shows a part of the numerical value graph for object $stus$ in our running example (where some data fields have been omitted for readability). A rectangle (with round corners) represents a categorical type, whereas a circle associated with the type denotes a numerical value which can be extracted from the type. For readability, each edge is labeled with an abbreviated variable name and each node is labeled with the type. Notice that a categorical type is always associated with a $boolean$ type value which is true iff the object is null. An edge reads as ``contains''.  For instance, an object of type $Stu[]$ contains objects of type $Stu$, which in turn contains three numerical-type variables $score$, $ID$ and $newscore$. In addition, each categorical type object is associated with a set of features which are the results of the inspector methods in the respective class, e.g., the returned value of $isEmpty()$ or $length()$ for a $String$ object.

Given a program state, we can build the numerical value graph of each variable and obtain a vector of features (i.e., the numerical values in the graph) systematically. One complication is that in order to apply classification techniques, each feature vector must have the same number of features. Different program states however may have different structures (e.g., two $String$ objects with different length) and therefore there are different numbers of features. In this work, we only use features which are common to program states in $O^-$ and $O^+$, e.g., for arrays with different sizes, we use features like its size, the value of the first/last element, etc. The underlying assumption is that these common features are sufficient to capture their difference. By focusing on the common features, we make sure the feature vectors are of the same size.

Another challenge is that there may be a large number of features and identifying the relevant features for generating the predicate is essential in our approach. This is a well-known problem for machine learning~\cite{DBLP:journals/jmlr/GuyonE03,DBLP:conf/icml/YangP97} as well as applications of learning techniques in software engineering community~\cite{Akalya:2012:SFP:2385865,Subramanian:1993:DRS:153678.153689,DBLP:journals/tse/ShepperdS97,DBLP:journals/tse/KeungKJ08}. In this work, we solve the problem heuristically by prioritizing the features based on the following two assumptions. First, we assume the recently accessed (read or written) features are more likely to be relevant. For instance, if we are to generate a classifier before line 8 in our running example, $score$ of a student object is considered relevant since it is accessed at line 7. Intuitively, this is because since the bug is likely at a previous location, the features accessed recently are likely useful in explaining the bug. Thus, we sort all the features according to when they are accessed (i.e., the more recent, the higher priority). Second, we assume that the features at the top of the numerical value graphes are more likely to be relevant. Intuitively, this is because those values are easier to access and thus are more likely to be relevant to the program behavior. Thus, we further sort the features so that if two features are both not accessed recently, the one near the top of the numerical value graph has the higher priority. For instance, given the $stus$ object in test case 1, the level 1 features (i.e., $[3,0,0,0]$ where the first number 3 means that the length of the array is 3 and the rest of the 0s mean that all $Stu$ objects in the array are not null) would have higher priority than the students' IDs according to this assumption.

Furthermore, because we prefer simple bug explanations, \textsc{Ziyuan} always attempts to generate a bug explanation using fewer features, i.e., starting with one feature with top priority for classification and gradually increasing the number if necessary. That is, \textsc{Ziyuan} starts by finding a classifier with the top feature; and then with the second top feature; etc., before trying to find a classifier with two or more features. For instance, in our running example, \textsc{Ziyuan} tries to identify a classifier based on $max$'s value only first; then a combination of $max$'s value and a feature of $stus$; and so on.

We acknowledge that the features obtained this way may not always be the \emph{best} to explain the bug. For instance, in our running example, a useful feature for explaining the bug would be the maximum score of all students, with which we can explain the bug as: the program is buggy because $max$ is not equal to the actual maximum at line 8. Nonetheless, our empirical study shows that features obtained using the above heuristics are often be useful in explaining the bug. We plan in future work to explore alternative ways of identifying relevant features (like Delta Debugging~\cite{DBLP:conf/esec/Zeller99} or feature selection methods used in machine learning~\cite{DBLP:journals/jmlr/GuyonE03,DBLP:conf/icml/YangP97}). \\

\noindent \emph{Example} In our running example, given the program location right after the first loop, there are two variables: $max$ and $stus$. Both of them are relevant (i.e., there is a dependency from the assertion to these variables). Next, since $max$ is accessed last, it has the top priority, followed by the features of $stus$. Of all the features of $stus$, feature $score$
has higher priority since it is accessed recently in the loop. Afterwards, the top level feature on whether it is null has the higher priority than the level 1 features, and then level 2 ones. Table~\ref{testcases} column 4 shows the level 2 features of $stus$, with the value for $max$, for each test cases.

\subsection{Step 4: Artificial Data Synthesis} \label{datasynthesis}
After the last step, we have transformed $O^+$ and $O^-$ into two sets of feature vectors, denoted as $F^+$ and $F^-$ hereafter. We then apply Support Vector Machines (SVM) to identify a predicate capturing the difference between $F^+$ and $F^-$. In order to learn an accurate classifier, a large number of samples (i.e., $F^+$ and $F^-$ in our setting), are required. A limited set of samples might result in a meaningless classifier. For instance, given the data in Table~\ref{testcases}, if we use $max$'s value to identify a classifier right before line 8, the result is: $-1*max \geq -92$. It translates: if $max < 92$ is satisfied, there is no assertion failure. It is obviously incorrect and the reason is the lack of sufficient test cases. In practice, we often have a limited set of user-provided test cases. In this work, we develop an approach to overcome the problem. One approach contains two parts. One is artificial data synthesis (this step) and the other is selective sampling (step 6). In the following, we explain how artificial data synthesis works. For simplicity, we focus on learning a classifier right before line 8 in our running example.

Given $F^+$ and $F^-$, we collect all possible values of each selected feature from $F^+$ and $F^-$. Next, for each value combination of the selected features, we mutate the program by adding a statement at the program location to set the respective variables to those values. For instance, if the selected feature is $max$'s value, based on the test cases shown in Table~\ref{testcases}, the possible values of $max$ are 94, 90 and 12. We mutate the given program into three different versions, one by adding a line before line 8 to set $max$ to 94; one by setting $max$ to 90; one by setting $max$ to 12. This is illustrated in Figure~\ref{illustrative}. Afterwards, we re-run the three test cases, for each mutated program and obtain the testing results. For instance, the additional testing results for our running example are shown in Table~\ref{mutestcases}, where the first row reads: setting $max$ to 90 right before line 8 and then running test 1 results in assertion failure. 
We remark that we re-run all test cases because we only set the value of some (not all) variables. Lastly, we update $F^+$ and $F^-$ based on the testing results, e.g., the feature at the first row of Table~\ref{mutestcases} is added into $F^-$ since the testing result is failure.

The benefit of the data synthesis is that we would have additional samples. For instance, with the additional data in Table~\ref{mutestcases}, $-1*max \geq -92$ is no longer a classifier since there are both passed and failed test cases with $max=94$. \emph{We remark that some of the feature vectors obtained this way at the given program location are not feasible in actual execution}. For instance, there is no test case which would reach the program point with the feature vector $[12,100,60,94,1,0,2,0,3,0]$ where $max = 12$ since 12 is not a score of any student. As a result, we would learn an over-approximation of the actual invariant (since it includes program states which are infeasible in the actual program). The additional samples are however helpful in pruning meaningless classifiers.

\begin{figure}[t]
\centering
\includegraphics[width=\linewidth]{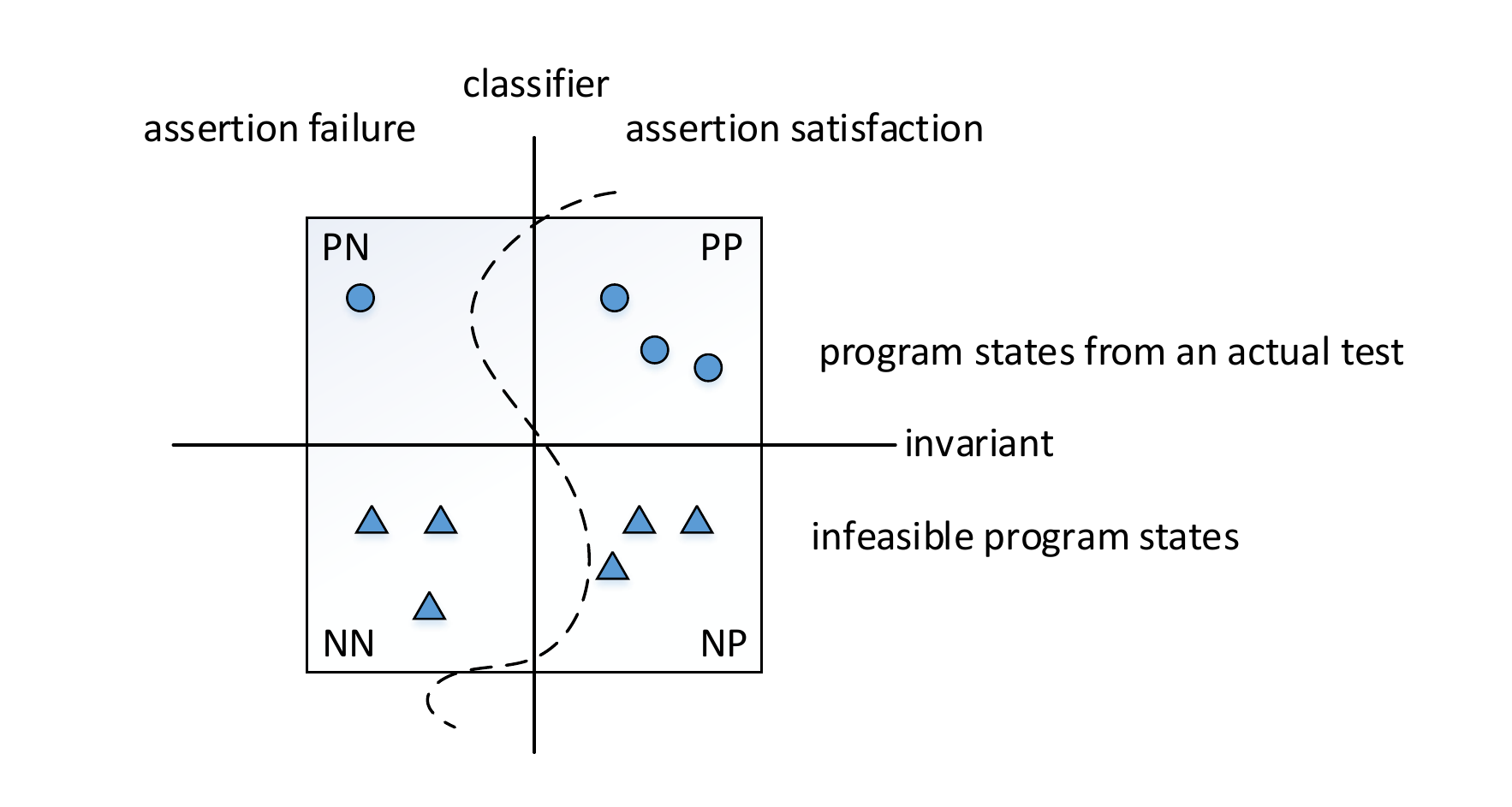}
\vspace{-8mm}
\caption{Classifier vs. Invariant}
\label{fig:tlv}
\end{figure}

Figure~\ref{fig:tlv} illustrates the categorization of the program states that we are getting through testing and data synthesis. It also shows the relation between the classifier that we are learning and the actual ``invariant'' at the program location. The circles represent the program states we obtain from the test cases (as in the running example) and the triangles represent the synthesized ones. The dashed line is to be ignored for now. There are four categories of program states: on the upper-right, we have those that lead to no assertion failure and can be obtained from an actual test (labeled $PP$); in the bottom-left, we have those that lead to assertion failure and cannot be obtained from any actual test (labeled $NN$); and the other two (labeled $PN$ and $NP$ respectively). Ideally, we should rely only on program states which can be obtained from actual test cases, i.e., the upper half of the space, and we would learn $classifer \land invariant$. The problem is we have a limited set of test cases, in particular, we often have very few failed test cases, and as a result, the classifier would be in-accurate. By using those program states obtained from testing results on the mutated programs, we would obtain program states not only in the upper half but also the bottom half, and therefore likely a more accurate classifier, as we have witnessed in our running example. This way, the classifier we obtain would be $classifier$, which is an over-approximation of $classifer \land invariant$. From another point of view, the program from the program point we are investigating to the assertion is never mutated. If we take that part of the program as a function, we are feeding arbitrary inputs to that function and the classifier is a predicate on the inputs which tells whether the function would output assertion failure or not.

\begin{table}[t]
\scriptsize
\centering
\caption{Testing results on mutated programs}
\begin{tabular}{|c|c|c|c|}
\hline
\textbf{test} & \textbf{mutation} & \textbf{Pass/Fail} & \textbf{ranked features} \\
\hline
1 & $max = 90$ & Fail & [90,100,60,94,1,0,2,0,3,0] \\
\hline
1 & $max = 12$ & Fail & [12,100,60,94,1,0,2,0,3,0] \\
\hline
2 & $max = 94$ & Pass & [94,75,90,80,3,0,2,0,1,0]\\
\hline
2 & $max = 12$ & Fail & [12,3,75,0,2,90,0,1,80,0] \\
\hline
4 & $max = 94$ & Pass & [94,-33,12,0,99,0,-10,0,0,0]\\
\hline
4 & $max = 90$ & Pass & [80,-33,12,0,99,0,-10,0,0,0] \\
\hline
\end{tabular}
\label{mutestcases}
\end{table}

\subsection{Step 5: Classification}
In the following, we present how we obtain a classifier automatically based on SVM. Given two sets of feature vectors $F^+$ and $F^-$, we apply an SVM-based approach to identify a classifier between them systematically. SVM is a supervised machine learning algorithm for classification and regression analysis. We use its binary classification functionality. Mathematically, the binary classification functionality of SVM works as follows. Given $F^+$ and $F^-$, it tries to find a half space $\Sigma_{i=1}^{n} c_i*x_i \geq c$ (where $c_i$ and $c$ are constant coefficients and $x_i$ are variables) such that (1) for every feature vector $[p_1, p_2,\cdots,p_n] \in F^+$ such that $\Sigma_{i=1}^{n}c_i*p_i \geq c$ and (2) for every feature vector $[m_1, m_2,\cdots,m_n] \in F^-$ such that $\Sigma_{i=1}^{n}c_i*m_i < c$. If $F^+$ and $F^-$ are linearly separable, SVM is guaranteed to find a half space. Furthermore, there are usually multiple half spaces that can separate $F^+$ from $F^-$. In this work, we always choose the \textit{optimal margin classifier} (see the definition in~\cite{Sharma2012}) if possible. This half space could be seen as the strongest witness why $F^+$ and $F^-$ are different. If, however, $F^+$ and $F^-$ cannot be perfectly classified by one half space only, we need to identify multiple half spaces, which together classifies $F^+$ and $F^-$. In the following, we present the classification algorithm we adopt~\cite{Sharma2012} for our task. We remark that \textsc{Ziyuan} is extensible so that different classification algorithms can be adopted.

\begin{algorithm}[t]
\SetAlgoVlined
\Indm
\KwOut{a classifier for $F^+$ and $F^-$}
\Indp
\If{SVM finds a classifier $f$ for $F^+$ and $F^-$} {
    return $f$;
}
let $hy$ = true\;
\While{$F^-$ is not empty} {
    remove an arbitrary $p$ from $F^-$\;
    \If{SVM finds a classifier $newhy$ for $F^+$ and $\{p\}$} {
        remove all $p' \in F^-$ such that $newhy(p') < 0$\;
        set $hy$ to be $hy \land newhy$\;
    }
    \Else{
        return null\;
    }
}

\Return $hy$\;
\caption{Algorithm $svm(F^+, F^-)$}
\label{alg:veri}
\end{algorithm}

Algorithm~\ref{alg:veri} shows the classification algorithm which is designed to find classifiers in the form of conjunctive linear inequalities. The inputs are $F^+$ and $F^-$. At line 1, we apply SVM to check whether there is a single half space which perfectly classifies $F^+$ and $F^-$. If there is, we return the classifier at line 2. Otherwise, it tries to identify a classifier composed of multiple half spaces. In particular, at line 3, variable $hy$ is declared which would hold the resultant classifier. The loop from line 4 to 10 then applies SVM multiple times to identify one half space at a time. At line 5, one arbitrary feature vector $p$ is picked from $F^-$. A new half space $newhy$ is then identified using SVM to classify $F^+$ and $\{p\}$. Next, all feature vectors in $F^-$ which are correctly classified using $newhy$ are removed from $F^-$. The loop terminates when every feature vector is correctly classified. It is not hard to see that this algorithm always terminates since $F^-$ is finite and its size decreases monotonically. \\

\noindent \emph{Example} In the following, we demonstrate how the classification works using our running example. For example, if we select $max$'s value and whether $stus$ is null or not as the relevant features, with the data in Table~\ref{testcases} and Table~\ref{mutestcases}, Algorithm~\ref{alg:veri} finds no classifier, since $[90,0]$ is both in $F^+$ and $F^-$. For the same reason, we would find no classifier if we use only $stus$' level-1 features or only one feature from all those level-2 features, e.g., only $stus[0].ID$, or only $stus[0].score$, etc. However, if we use the following two features: $max$'s value and $stus[0].ID$, we obtain the classifier: $0.026*max + 0.980*stus[0].ID \geq 4.305$ where the numbers are rounded off to three decimal places for simplicity. The classifier is correct with respect to all the data we have in Table~\ref{testcases} and Table~\ref{mutestcases}. It is however not meaningful. In the following, we discuss how to fix this problem.

\subsection{Step 6: Selective Sampling}
The above example shows that, even with artificial data synthesis as discussed in Section~\ref{datasynthesis}, the classifier might still be incorrect, due to the lack of samples. In fact, without feature vectors right by the `actual' classifier, it is very unlikely that we would find the actual classifier. For instance, if the actual classifier is $x \geq 33$ and $F^+$ contains only $x=100$ and $F^-$ contains only $x=0$, we are unlikely to generate the classifier $x \geq 33$. However, if $F^+$ contains samples right on or nearby the actual classifier, e.g., $F^+$ contains $x=33$ and $F^-$ contains $x=32$, it is more likely that SVM will generate the classifier $x \geq 33$.

This is illustrated at a high level in Figure~\ref{fig:tlv}. There could be many classifiers separating those samples in $PP$ and $NP$ from those in $PN$ and $NN$. The dashed line presents one example of them. Intuitively, in order to get the `actual' classifier, we need samples which would distinguish the actual one from any nearby one. This problem has been addressed in the machine learning community through active learning and selective sampling~\cite{DBLP:conf/icml/SchohnC00}. The idea is to repeatedly generate samples nearby the current classifier and then re-classify to identify an improved classifier. In particular,
SVM selective sampling techniques have been shown to identify accurate classifiers through the process in many applications~\cite{DBLP:conf/mm/TongC01,DBLP:journals/jmlr/TongK01}. In the following, we skip details on selective sampling and focus on how it is applied in our work. To the best of our knowledge, ours is the first technique applying selective sampling to solve a software engineering problem.

Algorithm~\ref{alg:sampling} presents details on how selective sampling is adopted in \textsc{Ziyuan}. At line 1, we obtain a classifier based on Algorithm~\ref{alg:veri}, which is in the form of a conjunction of multiple half spaces. We then apply selective sampling to compute feature vectors which are close to the classification boundary (a.k.a.~the most informative samples). In particular, at line 5, we apply standard techniques~\cite{DBLP:conf/icml/SchohnC00} to identify two points on the boundary of each half space. Next, for each computed point (i.e., a feature vector), right before the program location, we mutate the program state according to the feature vector. Following the above simple example, if $F^+$ contains only $x=100$ and $F^-$ contains only $x=0$ initially, we get the classifier $x \geq 50$. Next, applying selective sampling, we obtain a point $x=50$ based on this classifier. Afterwards, at the program location where we are learning, we add a statement assigning $x$ to 50. The intuition is that if the classifier $x \geq 50$ is correct, once we assign $x$ to 50 there, there should not be assertion failure any more. Afterwards, we execute the test cases and update $F^+$ and $F^-$ accordingly at line 9 and 11, based on the testing results. For instance, in the above example, since we assume the actual classifier is $x \geq 33$, executing the test case would not lead to assertion failure and therefore $x=50$ will be added into $F^+$. We then apply Algorithm~\ref{alg:veri} again to get a new classifier at line 12. If the newly identified classifier differs from the old one, we repeat the process; otherwise we return the newly identified classifier. For instance, following the above example, since $F^+$ now contains $x=50$ and $x=100$, and $F^-$ contains $x=0$, the new classifier is $x \geq 25$. Since it is different from $x \geq 50$, we repeat the process. Since the classifier is improved upon every iteration, it converges to an accurate one. In the above example, we would subsequently find the classifier $x \geq 38$ (after rounding up), then $x \geq 32$, then $x \geq 35$, then $x \geq 34$ and finally $x \geq 33$. \\

\noindent \emph{Example} As presented above, in our running example, due to the very limited set of test cases, the first classifier using $max$'s value and $stus[0].ID$ is $0.026*max + 0.980*stus[0].ID \geq 4.305$, i.e., the value of $clf$ at line 1 of Algorithm~\ref{alg:sampling}. At line 5, we obtain the following most informative samples: $[90,2]$ and $[128,1]$. They are obtained by taking existing feature values and solve for the other based on the current classifier. That is, we take $max$ to be 90 and solve $0.026*max + 0.980*stus[0].ID = 4.305$ and get $stus[0].ID = 2$. Similarly, we get the other pair by taking $stus[0].ID$ to be 1. Next, we mutate the program by inserting $max = 90$ and $stus[0].ID = 2$ right before line 8 in the program. We re-run the three test cases and we obtain the additional samples in Table~\ref{mutestcases2}. Next, at line 12, invoking Algorithm~\ref{alg:veri} returns $null$ since $[90,2]$ is both labeled in $F^+$ and $F^-$ (i.e., the same feature vectors are both positive and negative). The algorithm then returns null at line 4 in the next iteration.

Next, \textsc{Ziyuan} tries to learn classifier with other features. For the same reason, \textsc{Ziyuan} finds that there is no classifier using features like $max$'s value with value of $stus[0].score$ (or $stus[1].score$). However, if we use $max$'s value and $stus[2].score$ as the relevant features, with only the data in Table~\ref{testcases},~\ref{mutestcases} and~\ref{mutestcases2}, the following is obtained before selective sampling is applied: $0.053*max-0.125*stus[2].score >= -6.058$.
Next, we apply selective sampling and keep computing new samples. For instance, one new sample is [74,80] (where $max$ is 74 and $stus[2].score$ is 80). After testing, it is added into $F^-$. Adding the new labeled samples, we obtain a better divider. After multiple iterations, the algorithm terminates and reports the classifier: $2*max-2*stus[2].score \geq -1$. Since both variables are integers, it is simplified as $max \geq stus[2].score$.

How do we interpret this result? Intuitively, what we learned is: assertion failure occurs if $max \geq stus[2].score$ is not satisfied. Thus, in order to make sure the assertion is always satisfied, the programmer should examine the predicate and decide whether it should be an invariant at the location. If it is, the program before the program location should be modified such that the predicate is always satisfied. For instance, for our running example, $max \geq stus[2].score$ should be an invariant and in this case it correctly suggests that $max$ is computed wrongly and therefore the program before line 8 must be modified. If the programmer decides that the predicate is not supposed to be an invariant, the program after the program location needs to be modified such that when the predicate is not satisfied, the assertion could still be satisfied. That is, the (negation of the) predicate captures a generalized case which is either not handled at all or not handled correctly in the program.

\begin{table}[t]
\scriptsize
\centering
\caption{Testing results on selective samples}
\begin{tabular}{|c|c|c|c|}
\hline
\textbf{test} & \textbf{mutation} & \textbf{Pass/Fail} & \textbf{L2 features} \\
\hline
1 & $[90,2]$ & Fail & [90,2,100,0,2,60,0,3,94,0] \\
\hline
1 & $[128,1]$ & Pass & [128,1,100,0,2,60,0,3,94,0] \\
\hline
2 & $[90,2]$ & Pass & [90,2,75,0,2,90,0,1,80,0]\\
\hline
2 & $[128,1]$ & Pass & [128,1,75,0,2,90,0,1,80,0] \\
\hline
3 & $[90,2]$ & Pass & [90,2,-33,0,-10,12,0,0,0,0]\\
\hline
3 & $[128,1]$ & Pass & [128,1,-33,0,-10,12,0,0,0,0] \\
\hline
\end{tabular}
\label{mutestcases2}
\end{table}

\begin{algorithm}[t]
\SetAlgoVlined
\Indm
\KwIn{$F^+$ and $F^-$}
\KwOut{a classifier for $F^+$ and $F^-$}
\Indp
let $clf$ = $svm(F^+, F^-)$\;
\While{true} {
    \Return $null$ if $clf$ is null\;
    compute the next sample $sam$ using selective sampling\;
    mutate the program according to $sam$\;
    \For{each test case going through the location} {
        \If{$test$ fails the assertion} {
            extract $f^-$ and add $f^-$ into $F^-$\;
        }
        \Else {
            extract $f^+$ and add $f^+$ into $F^+$\;
        }
    }
    $newclf$ = $svm(F^+, F^-)$\;
    \If{$newclf$ differs from $clf$} {
        $clf$ = $newclf$\;
    }
    \Else {
        \Return $newclf$\;
    }
}
\caption{Algorithm $classify(F^+, F^-)$}
\label{alg:sampling}
\end{algorithm}

\subsection{The Overall Algorithm} \label{overall}
We are now ready to present the overall approach of \textsc{Ziyuan}, which is shown in Algorithm~\ref{alg:veri2}. \textsc{Ziyuan} has four configurable parameters. $M$ is the number of random test cases to be generated; $X$ is a threshold on the suspiciousness score, i.e., only those program locations with a suspiciousness more than $X$ are examined; $N$ is the maximum size of the feature vectors; and $K$ is the maximum number of features used in a classifier. We start with generating $M$ random test cases and categorize them into failed ones and passed ones. Next, we apply bug localization to identify a list of program locations to generate likely invariants. For each program location with suspiciousness more than $X$, we identity two set of ordered feature vectors. For each combination of $K$ or less features out of a total of $N$ features, we apply artificial data synthesis and classification and selective sampling, to search for a classifier. Anytime a classifier is identified, we terminate and report it as the bug explanation. Note that it may find a classifier composed of many half spaces, which could be complicated for user comprehension. Thus, we throw away the classifier if it contains more than a threshold number of (3 by default) half spaces. The algorithm terminates when we exhaust the program locations and features.

The classifier identified by the algorithm is always correct with respects to the feature vectors (which are either obtained through the test cases or synthesized in the process). Since there are only finitely many combinations of program locations and features, Algorithm~\ref{alg:veri2} is always terminating. We roughly measure the complexity of the algorithm in term of the number of calls of the SVM classification algorithm. It is bounded by $\#X*C_{N+K-1}^K$ where $\#X$ is the number of program locations with suspiciousness more than $X$ and $C_{N+K-1}^K$ is an upper bound for $C_N^1 + C_N^2 + \cdots + C_N^K$. In practice, $\#X$ is often limited to be a small number like 10 (i.e., we examine the top 10 bug locations (after grouping consecutive ones) and $K$ is 3 by default and $N$ is 10 by default. As a result, the above complexity is often manageable in our experiments. 
\section{Implementation and Evaluation}\label{experiments}
Our approach has been implemented as a toolkit named \textsc{Ziyuan} (available at~\cite{ziyuandata}). \textsc{Ziyuan} is built upon a number of open source software projects, including (1) a re-implementation of the Randoop algorithm, extending~\cite{randooptool} with support for Java interfaces; (2) Javaslicer~\cite{jslicer} for dynamic program slicing; (3) the JaCoCo Java code coverage library~\cite{jacoco} for collecting code coverage information; (4) the LIBSVM library for SVM~\cite{libsvm}; and Java ILP, a Java interface to ILP solvers~\cite{ilp}, which is used for selective sampling. In the following, we evaluate \textsc{Ziyuan} in order to answer three research questions (RQ).

\begin{algorithm}[t]
\SetAlgoVlined
\Indm
\KwIn{program $Prog$; failed test set $F$ and passed test set $P$}
\KwOut{a likely invariant}
\Indp
generate $M$ random test cases\;
execute them and add them to $F$ and $P$ accordingly\;
identify a list of $X$ potential bug locations\;
\For{each bug location $b$ in the list} {
    extract a set of $N$-dimension feature vectors $F^+$ from $P$\;
    extract a set of $N$-dimension feature vectors $F^-$ from $F$\;
    \While{there is a new combination}{
        select a a combination of $K$ or less out of $N$ features\;
        apply artificial data synthesis and update $F^-$ and $F^+$\;
        let $exp = classify(F^+, F^-)$\;
        \If{$exp$ is not $null$ and contains 3 clauses or less} {
            \Return the classifier\;
        }
    }
}
output "no explanation is identified"\;
\caption{The Overall Algorithm: $explain(Prog, F, P)$; ;  }
\label{alg:veri2}
\end{algorithm}

Our test subjects include 21 real-world bugs from open source projects including the JavaParser1.5 project (JP), the Java-diff-utils project (JDU), the Joda-Time project (JT) and Apache Commons Math library (ACM)), from the bug collection in~\cite{JustJE2014} (D4J) and the bugs discovered in~\cite{DBLP:conf/kbse/ZhangZE11}. These bugs are selected based on the following criteria. First, we select bugs which are relatively easier to understand. This is because we aim to manually specify the initial assertion as well as to check whether the generated predicate is relevant. Second, we select those buggy programs with at least one passed test case. Lastly, we are limited to buggy programs which do not rely on Java features which are not yet supported in \textsc{Ziyuan} (e.g., abstract methods). The bugs are summarized in Table~\ref{bugtable}, where the first column shows the project name, the second column shows the issue number and the third column is the link to the bug report. Note that a `-' in the table means the information is skipped as it is irrelevant or not available.

For each bug, we manually created an initial assertion according to the bug report. This is often straightforward if the bug results in an exception, i.e., we find the line where the exception is thrown and add an assertion to turn the exception into assertion failure. For the sake of repeatable experiments, we disable random test generation for all the experiments (i.e., set $M$ to be 0) and use only existing test cases in the projects with an additional failed test case created according to the bug report. In general \textsc{Ziyuan} works better with more test cases. Notice that we manually remove the assertions in the test cases so that a test case fails if and only if the assertion in the program is violated. Furthermore, we set \textsc{Ziyuan} to focus on program locations with a suspicious score of 0.5 or above. \textsc{Ziyuan} is set to search for a classifier constituted by at most 3 features from the top 10 features. Lastly, SVM often takes a long time if there is no linear classifier and therefore we set a 5 second time out for each invocation of SVM. Details of the projects and the bugs, along with our analysis logs can be found at~\cite{ziyuandata}. \\

\begin{table*}[t]
	\caption{Efficiency evaluation}
	\label{bugtable}
\begin{center}
\begin{tabular}{|c|c|c|c|c|c|c|c|c|}
\hline
Project & Issue \# & URL & LOCfail & Time & Relevance & Daikon & Ochiai vs. \textsc{Ziyuan} \\
\hline
JP & 46 & \cite{bug1} & 707 & 3m & Missing Case & to & 29/\textbf{3} \\
JP & 57 & \cite{bug2}& 1154 & 15m & Invariant & to & 48/\textbf{39} \\
JDU & 10 & \cite{bug3}& 85 & 73s & Invariant & $\times$ & 81/\textbf{6} \\
JT & 227 & \cite{bug4}& 1109 & 4m & Incorrectly Handled Case & error & \textbf{3}/55 \\
JT & 21 & \cite{bug5}& 1113 & 24s & Incorrectly Handled Case & error & 43/\textbf{2} \\
JT & 77 & \cite{bug6}& 1210 & 61s & Missing Case & error & 54/\textbf{15} \\
ACM & 835 & \cite{bug7}& 18 & 7m & Invariant & $\times$ & \textbf{2}/3 \\
ACM & 1196 & \cite{bug8}& 152 & 42s & Incorrectly Handled Case & error & 152/\textbf{1} \\
ACM & 1005 & \cite{bug9}& 19 & 4m & Invariant & error & 4/\textbf{1} \\
D4J Time & 8 & \cite{JustJE2014} & 5 & 69s & Incorrectly Handled Case & error & 2/\textbf{1} \\
D4J Math & 1,4,38,40,58,61,70,79,84 & \cite{JustJE2014} & - & 13m(total) & Inconclusive & - & - \\
FailureDoc 1 & - & \cite{bug11} & 576 & 33s & Incorrectly Handled Case & + & - \\
FailureDoc 2 & - & \cite{bug11} & 64 & 75s & Missing Case & + & - \\
\hline
\end{tabular}
\end{center}
\vspace{-4mm}
\end{table*}


\noindent \textbf{RQ1: Is \textsc{Ziyuan} sufficiently efficient?} We first evaluate whether \textsc{Ziyuan} is sufficiently efficient for practical usage. The fifth column of Table~\ref{bugtable} shows the average execution time of \textsc{Ziyuan} over 10 executions for each bug. The experiments were conducted in Windows 7 on a machine with an Intel(R) Core(TM) i5-2430m, running with one 2.40GHz CPU, 4M cache and 8 GB RAM. The data shows that \textsc{Ziyuan} takes a few minutes to generate the predicates, which we believe is reasonably efficient, since it usually takes hours to fix a bug~\cite{DBLP:conf/msr/KimW06}. To show that these bugs are not trivial (e.g., it is hard to trace the failed test case step-by-step to locate the bug), the 4th column shows the number of statements executed in the failed test case (excluding external library calls). Though some bugs have relatively few number of statements, they often rely heavily on external library calls.

We remark that sound optimization have been implemented in \textsc{Ziyuan} to improve its efficiency. For instance, Algorithm~\ref{alg:sampling} may take many iterations to converge. In order to reduce the number of iterations, each time a classifier is identified, we make use of the type information for better selective sampling. For instance, after calculating a new sample $[x,y]$ with two integer-type features at line 4 of Algorithm~\ref{alg:sampling}, we additionally check and label nearby samples, for instance $[x+1,y],[x,y+1],[x-1,y],[x,y-1]$, so that Algorithm~\ref{alg:sampling} converges fast. \\

\noindent \textbf{RQ2: Does \textsc{Ziyuan} generate useful bug explanations?} We acknowledge that it is subjective on whether a predicate learned by \textsc{Ziyuan} is useful in explaining the bug. In the following, we attempt to answer this question in three ways. First, we check whether the predicate is relevant by manually examining the corresponding bug fixes. Second, we present specific findings for some of the bugs and the reason why we believe the bug explanation is useful, so that the readers can judge by themselves. Third, we conduct a user study to see whether the bug explanations are useful for bug understanding and fixing. We present the details below. \\

\noindent \emph{Relevance} Recall that a predicate generated by \textsc{Ziyuan} could be either an actual invariant (which is violated due to a bug) or a predicate that captures a generalized case which is not handled at all (i.e., a missing case) or handled incorrectly. Thus, if the generated predicate is `correct', either the bug should be fixed such that the predicate becomes an invariant or specific code is introduced to handle the case when the predicate is not satisfied. We manually examine the bug fixes to check whether it is the case for each bug. If the answer is yes, we consider that the predicate is relevant. Notice that some of the bugs were open and thus we proposed the fixes based on our analysis and confirmed them with the authors.

The results are summarized in Table~\ref{bugtable} column ``Relevance''. For all bugs, the predicate generated by \textsc{Ziyuan} is satisfied in all the passed test cases and is not satisfied in the failed test case. Note that for 9 bugs in ACM, due to our limited understanding of ACM's implementation, we are not yet to be able to confirm whether the generated predicate is related to the actual cause of the bug. For the rest, in 4 cases, the fixes precisely make the learned predicate an invariant at the program location. In 3 cases, the program is fixed by introducing code to handle the case when the learned predicate is not satisfied. In 5 cases, the program is modified so that it handles the case when the learned predicate is not satisfied differently. We conclude that the predicates are relevant in these 12 cases. \\

\noindent \emph{Specific Findings} Next, we present sample findings of the bugs and the generated predicates.

The JP project aims to build a Java 1.5 parser with AST generation and visitor support. The AST records the source code structure, javadoc and comments; and supports changing the AST nodes or creating new ones. 
\textsc{Ziyuan} is applied to analyze an open bug (issue 46) and a closed bug (issue 57) for this project.

The bug report for issue 46 contains the following information. After parsing the Java program shown below, the output of the method \emph{CompilationUnit.toString()} in JavaParser1.5 prints only comment 3, whereas it should print all three comments.

{\scriptsize \begin{verbatim}
    /** Comment 1*/
    /** Comment 2*/
    /** Comment 3*/
    package net.perfectbug.test;
    public class Test {}
\end{verbatim}}
With the information, we first manually created a test case according to the report. Next, we added an assertion in JavaParser1.5 to assert that after parsing the above program, invoking \emph{CompilationUnit.getComments().size()} would return more than 1 (i.e., there should be more than 1 line of comments). 
We then fed the program, the failed test case, along with existing passed test cases to \textsc{Ziyuan}. After program slicing, testing and learning, tracking through 7 classes, \textsc{Ziyuan} outputs a message which says that the assertion is satisfied if $special.specialToken.isNull$ is true at line 67 of class $japa.parser.ASTParserTokenManager$; otherwise, it fails.

\begin{figure}[t]
{\scriptsize \begin{verbatim}
57. private void CommonTokenAction(Token token) {
58.   lastjavadoc = null;
59.   if (token.specialToken != null) {
60.      if (comments == null) {
61.         comments = new LinkedList<Comment>();
62.      }
63.      Token special = token.specialToken;
64.      if (special.kind = JAVA_DOC_COMMENT) {
65.         lastJavaDoc = ...;
66.         comments.add(lastJavaDoc);
67.      } else if (special.kind==SINGLE_LINE_COMMENT) {
68.         LineComment comment = ...;
69.         comments.add(comment);
70.      } else if (special.kind==MULTI_LINE_COMMENT) {
71.         BlockComment comment = ...;
72.         comments.add(comment);
73.      }
74.   }
75. }
\end{verbatim}}
\vspace{-5mm}
\caption{Sample code from JavaParser1.5}
\label{code}
\end{figure}

Without knowing how JavaParser1.5 is implemented, we examine the code around line 67, as shown in Figure~\ref{code}. By checking the value of $special.specialToken.isNull$ in the test cases, we realize it is not true only if there are multiple consecutive comments before a token (which could be a class or statement). Furthermore, variable $comments$ contains only the last comment (not all comments) when $special.specialToken.isNull$ is not true, which according to \textsc{Ziyuan}, is when a test fails. Since $special.specialToken.isNull$ being true is not likely an invariant at this program location, we conclude that it signals a missing case, i.e., the authors forgot to handle the case when there are multiple consecutive comments. We then fixed the bug by introducing a while loop to add the multiple comments one-by-one if $special.specialToken.isNull$ is not true, replacing the block from line 59 to 74 in Figure~\ref{code}. The bug is then confirmed fixed (by the authors).

We also applied \textsc{Ziyuan} to issue 57 which reports that a particular method signature is parsed incorrectly. Without any knowledge on how the parsing works, we added a trivial assertion (without any generalization) to say that if the input is this particular method signature, the result should be certain particular string. \textsc{Ziyuan} identified a likely invariant: $type.typeArgs.isNull==true$, at line 1755 in class $ASTParser$, which reads: if $type.typeArgs.isNull$ is true, the failure does not occur. The actual fix (by the project authors) is at line 1810 (which is 4 statements before executing line 1755) and the fix is the insertion of the statement: $type.typeArgs = null$, which makes the learned predicate an invariant.

The two examples so far resulted in predicates constituted by boolean variables only. In the following, we show examples where selective sampling helps us to generate the exact boundary conditions. We applied \textsc{Ziyuan} to three issues in ACM: 835, 1196 and 1005. In particular, issue 1196 is a bug which is still open. It states that if variable $x$ is set to be \emph{0x1.fffffffffffffp-2} (equivalent to value 0.49999999999999994), $FastMath.round(x)$ returns 1 instead of 0 while clearly $x < 0.5$. We instrumented the program to assert that if a number is less than 0.5, the rounding result should be less than 1. \textsc{Ziyuan} tracked to the statement $x+0.5$ in the program and started finding classifiers. In our first attempt, \textsc{Ziyuan} failed to identify any classifier after a while. Our investigation shows that after a few iterations, the classifier becomes $x \leq 49991269898708784$, LIBSVM fails to classify the samples because the samples are too close. We then implemented a simple classification algorithm (and a simple solver for the same reason) to learn classifiers in the form of $x \geq c$ and obtained a predicate $x \leq 0.49999999999999991$. It means that when $x$ is smaller than the number, the rounding result is correct. This result in fact generalizes an open bug in JDK 6 and 7 (bug number JDK-6430675) by giving a range of $x$ which could trigger the bug. For issue 835, \textsc{Ziyuan} discovered that a likely invariant $fraction.numerator >= 0$ is violated in the failed test cases, which turned out to be the result of an integer overflow. A similar discover has been made for issue 1005.

We applied \textsc{Ziyuan} to analyze three issues of JT: 21, 27 and 227. Issue 227 reports that adding 50 days from May 15 results in June 4, which is clearly wrong. We added an assertion before method \emph{AddDays} in class \emph{MonthDay} and \textsc{Ziyuan} generated the predicate $days+iValues[1]\leq62$, which reads that if the number of days to be added plus the original day is larger than 62, the bug occurs. It points to a bug which is activated only if the resultant date is in the next-next month or later. Due to the space limit, we skip the details on \textsc{Ziyuan}'s findings for other bugs in the JT project or the JDU project. Interested readers are referred to~\cite{ziyuandata} for the details. Though limited in the number of test subjects, we confirm \textsc{Ziyuan} to be useful in helping users to understand these bugs. \\

\noindent \emph{User study} Finally, we perform a user study to evaluate whether independent programmers consider the generated predicates useful. The user study is conducted with 12 programmers (including PhD students, research assistants and research fellows). The programmers have a various number of years of programming experience (from 2 to 9 with an average of 5.75). They were divided into two groups randomly. The programmers in the first group were instructed to fix JP issue 46 without \textsc{Ziyuan}'s help and then to fix JDU issue 10 with \textsc{Ziyuan}'s help. The other group were instructed to fix the former issue with \textsc{Ziyuan}'s help and then the latter issue without \textsc{Ziyuan}'s help. This experiment is thus similar to a scenario where \textsc{Ziyuan} is used to help a programmer to fix a bug in the legacy code. These two bugs are chosen as they are representative. They are however not easy to fix.

Each programmer was given at most 30 minutes to study the bug so as to figure out precisely the reason of the bug and propose a fix if possible. We then evaluated whether their explanation and proposal were correct. The result is as follows. For the first bug, with \textsc{Ziyuan}'s help, 3 out of 6 programmers figured out the bug correctly in 10, 27, 30 minutes respectively. Without Ziyuan's help, 2 out of 6 did it in 14 and 30 minutes respectively. For the second bug, with \textsc{Ziyuan}'s help, 4 out of 6 programmers did it in 15, 23, 24 and 30 minutes respectively. Without Ziyuan, none of the programmers did it. Furthermore, all of the programmers agree that the information provided by \textsc{Ziyuan} was helpful. We take this as a positive feedback on the usefulness of the generated predicates. Note that the amount of time used by the programmers should be taken with a grain of salt as they often spend considerable time testing their bug understanding by trying to fix it, before reporting their finding. We acknowledge that the user study is limited in the number of programmers and bugs. We refer the readers to~\cite{ziyuandata} for the details on the user study. \\

\noindent \textbf{RQ3: Does \textsc{Ziyuan} complement existing approaches?} \textsc{Ziyuan} can be categorized as an invariant learning tool. Thus, we performed experiments to compare \textsc{Ziyuan} with the popular invariant generator \textsc{Daikon} as well as FailureDoc reported in~\cite{DBLP:conf/kbse/ZhangZE11}. To compare with \textsc{Daikon}, we use the same set of passed test cases used in \textsc{Ziyuan} for each project and check whether \textsc{Daikon} can learn an invariant which is relevant (as defined above). Note that \textsc{Daikon} does not learn from failed test cases. Furthermore, the `artificial' program states generated by \textsc{Ziyuan} do not constitute actual test cases and thus cannot be used by \textsc{Daikon} or FailureDoc. The results are summarized in column \textsc{Daikon} of Table~\ref{bugtable}, where $error$ means an exception; $to$ means timeout after one hour; $\times$ means none of the learned invariants are relevant and $+$ means some invariants are relevant. \textsc{Daikon} failed to learn useful invariants in most of the cases.

Similar to \textsc{Ziyuan}, FailureDoc aims to explain a failed test case. However, it focuses on the failed test case only (without analyzing the source code) and generates a predicate constituted by variables used in the failed test case only. In a way, it can be considered as applying \textsc{Ziyuan} with the following restrictions: (1) learning based on the variables in the failed test case only, using \textsc{Daikon} to generate a likely invariant, and not applying selective sampling. We tried FailureDoc on the list of bugs \textsc{Ziyuan} analyzed and had no useful results because FailureDoc does not support user-provided assertions. As shown above, we managed to apply \textsc{Ziyuan} to some of the bugs analyzed by FailureDoc in~\cite{DBLP:conf/kbse/ZhangZE11} and generated useful bug explanation in the program. We conclude that FailureDoc and \textsc{Ziyuan} are useful in different settings.

\textsc{Ziyuan} has a different goal from SBFL. However, we show that \textsc{Ziyuan} could potentially be used to improved SBFL. The last column of Table~\ref{bugtable} shows two numbers. The first one is how many statements must the user examine before reaching the statement containing the bug, assuming that the user examines the program statement-by-statement based on the suspiciousness ranking generated by Ochiai's approach. The second one is the number of statements the user has to examine, assuming the user starts with where \textsc{Ziyuan} generates the bug explanation and works towards the bug following the statements executed in the failed test case. A smaller number (highlighted in bold) is better since fewer statements are to be examined. Note that we do not have the fixes for the bugs presented in the last three rows and thus we skip them for this comparison.

Firstly, it can be observed from the data that SBFL may not always be effective, which is consistent with the observations in~\cite{DBLP:conf/issta/ParninO11,survey}. Second, though \textsc{Ziyuan} relies on bug localization, we observed in 8 out of 10 cases that the predicate is not generated at the most suspicious program location, but a program location closer to where the bug is in the code. Intuitively, this could be explained as follows: where the bug is easier to explain may also be where the fix is easier to fix. In the case of JT issue 227, the bug explanation is far from the bug because a large part of the relevant codes are a recursive method (i.e., method $add$ in class $BaseDateTimeField$) and \textsc{Ziyuan} currently tries to explain the bug only before or after loops or recursive methods. Though the number of bugs we studied is limited, the results suggest \textsc{Ziyuan} may complement SBFL. \\

\noindent \textbf{Limitations} \textsc{Ziyuan} has a number of limitations. First, though artificial data synthesis and selective sampling help to overcome the lack of test cases, the quality of the generated predicate may still depend on the test cases. For instance, in the extreme case, if no other test cases other than a failed test case is provided, neither artificial data synthesis nor selective sampling would help. To overcome this limitation, we are currently working on integrating \textsc{Ziyuan} with sophisticated testing engines to boost its performance.

Second, the effectiveness of \textsc{Ziyuan} relies on the user-provided assertion. In general, the stronger the initial assertion is, the stronger a bug explanation might be generated. For instance, in our running example, if we replace the assertion at line 3 with: $s1.newscore \leq 100 \land s2.newscore \leq 100 \land s3.newscore \leq 100$ (i.e., all students' new score must be no more than 100), the learned predicate is $stus[0].score \leq max \land stus[1].score \leq max \land stus[2].score \leq max$. We are currently investigating how to automatically generate the initial assertion.

Third, in general we cannot guarantee that the learned predicate is satisfied \emph{if and only if} the given assertion is satisfied. This problem can be solved by applying program verification techniques, i.e., to verify that the learned predicate is the weakest precondition of the program from the learning program location to the assertion, with respect to the assertion. Nonetheless, existing program verification techniques often have their own limitations and may not scale to complicated programs that we would like to handle.

Fourth, the effectiveness of \textsc{Ziyuan} depends on identifying the right features. Although our heuristics for feature selection worked in our empirical study, in general feature selection is challenging. We are investigating whether we can use advanced program analysis or feature selection methods to identify the relevant features automatically. Furthermore, \textsc{Ziyuan} currently does not use inspector method results other than those returning boolean values as features for learning. This is because, unlike instance variables which we can change their values during selective sampling, changing the returned values of inspector methods are challenging in general.

Fifth, the classification algorithm used in \textsc{Ziyuan} is limited to predicates in certain form. They may not be sufficient sometimes, e.g., the actual predicate could be non-linear or disjunctive. We are currently investigating different classification algorithms (e.g., SVM with kernel methods and neutral network) to overcome this problem. The challenge however is ensuring that the learned classifier is comprehensible by programmers.

Lastly, our empirical study is limited in the number of studied subjects and varieties. We are currently extending our collections of programs and bugs for further study.

\section{Conclusion and Related Work}\label{related}
The main contribution of \textsc{Ziyuan} is the application of invariant learning for bug explanation, as well as a novel approach to overcome the problem of lack of test cases in practice.
In essence, what \textsc{Ziyuan} does is to propagate the initial user-provided assertion through the program to a location that is close to where the bug is. We believe that this is useful as programmers could then compare our bug explanation with their understanding of the program specification. 

This work is also inspired by the line of work by Zeller and his collaborators, e.g.,~\cite{DBLP:conf/esec/Zeller99,DBLP:conf/sigsoft/Zeller02,DBLP:conf/icse/CleveZ05,DBLP:conf/issta/RobetalerFZO12,DBLP:journals/tse/GaleottiFMFZ15}. In particular, this work is closely related to the work in~\cite{DBLP:conf/issta/RobetalerFZO12}. In~\cite{DBLP:conf/issta/RobetalerFZO12}, the authors proposed to isolate bug causes through directing test case generation (based on~\cite{DBLP:conf/qsic/FraserA11}) towards certain factors which are potentially associated with the bug cause. Two kinds of factors are considered: the executed branches and state predicates. Similarly, \textsc{Ziyuan} identifies the bug causes in the form of state predicates. The state predicates used in~\cite{DBLP:conf/issta/RobetalerFZO12} (based on their previous work~\cite{DBLP:conf/issta/FraserZ11}) include comparison between accessible variable values at certain program locations, whereas \textsc{Ziyuan} relies on SVM to learn more complicated predicates. This work is related to previous work on using likely invariants for debugging~\cite{DBLP:conf/asplos/SahooCGA13,DBLP:conf/icse/HangalL02,DBLP:journals/corr/cs-SE-0310040}. Furthermore, this work is related to partial specification generation using symbolic methods~\cite{DBLP:journals/tosem/QiRLV12,DBLP:conf/cav/JoseM11,DBLP:conf/pldi/JoseM11}. \textsc{Ziyuan} complements the above work by using SVM to discover relevant state predicates and, novelly, a way of ``testing'' and refining the predicates (e.g., by selective sampling).

This work is inspired by the line of work on invariant learning by Ernest and his collaborators~\cite{DBLP:conf/icse/ErnstCGN99,DBLP:conf/sigsoft/NimmerE02,DBLP:conf/issta/NimmerE02,DBLP:conf/sigsoft/PerkinsE04,DBLP:conf/kbse/ZhangZE11}. In particular, this work is closely related to the work documented in~\cite{DBLP:conf/kbse/ZhangZE11}, which shares the same goal of explaining failed tests by inferring likely invariants. Their approach is to generate mutated tests based on the failed test case, obtain a set of failure-correcting objects and use \textsc{Daikon} to summarize properties of the failure-correcting objects, and lastly translate them into explanatory code comments. \textsc{Ziyuan} complements their work by analyzing not only the failed test case but also the code, and in the way how mutated tests are generated (e.g., selective sampling) and how the properties of the failure-correcting objects are generated.

This work is related to the work in~\cite{unpub}, where the authors learn a model in the form of finite state-automata to represent the scenarios in which errors occur. Our work has a different goal and a different learning approach. This work is related to work on explaining counterexamples, e.g.,~\cite{DBLP:journals/fmsd/BeerBCOT12} using the notion of causality and~\cite{DBLP:conf/spin/GroceV03} which is similar to delta debugging~\cite{DBLP:conf/esec/Zeller99}, and~\cite{DBLP:conf/pldi/LiblitAZJ03}. In contrast, we focus on learning a local invariant which helps bug understanding.

This work benefited from ideas from existing work on specification learning, including~\cite{tzuyu:ASE2013,fse,psyco:SAS2012,alur:POPL2005,sigmastar:POPL2013,psyco:SAS2012,xpsyco:ISSTA2013,DBLP:conf/sigsoft/BeschastnikhBSSE11,DBLP:journals/tse/CasoBGU12,DBLP:journals/tse/HughesB08,DBLP:conf/icse/CsallnerS06}. \textsc{Ziyuan} uses SVM-based learning to discover new predicates, which is similar to previous work in~\cite{tzuyu:ASE2013,fse}. In~\cite{tzuyu:ASE2013}, random testing and SVM are used to learn a typestate for Java classes. Later, the work in~\cite{fse} extends~\cite{tzuyu:ASE2013} to provide correctness and accuracy guarantee of the learned typestate. This work is different as we have a different objective (i.e., bug explanation) and a different learning approach, i.e., instead of L*~\cite{tzuyu:ASE2013,fse}, we use active learning and selective sampling for discovering invariants. This work is related to work on inferring documentation from programs as \textsc{Ziyuan} also learns program invariants. Examples include~\cite{DBLP:conf/sigdoc/RoachBT90} which facilitates programmers to write documentations,~\cite{DBLP:conf/issta/BuseW08a} which infers documentations from exceptions,~\cite{DBLP:conf/kbse/BuseW10} from software changes, etc. Our work is different as it is motivated for bug explanation.

In addition, this work is related to research on bug/fault localization, including but not limited to~\cite{survey,DBLP:conf/icst/MoonKKY14,DBLP:conf/issta/RobetalerFZO12,DBLP:conf/kbse/FleureyTB04,DBLP:journals/sqj/MisirliBT11}. Our work complements bug localization techniques by providing an explanation of the bug. Not only \textsc{Ziyuan} can benefit from better bug localization, but also the bug explanation identified by \textsc{Ziyuan} could potentially help pinpoint where the bug is. This work is broadly related to research on the art of debugging, e.g.,~\cite{DBLP:conf/oopsla/GuBSS12,debugging,DBLP:journals/tse/KimWKZCP11,DBLP:journals/corr/CornuBSM15}, as well as recent studies on program repair, e.g.,~\cite{DBLP:conf/icse/GouesDFW12,DBLP:conf/kbse/KeSGB15,DBLP:conf/popl/LongR16}.

\bibliographystyle{abbrv}
\bibliography{learning}
\end{document}